\documentclass[a4paper]{jpconf}
\usepackage{graphicx}
\begin{document}
\title{Parton density constraints from massive vector boson production at the LHC}

\author{Michael Klasen}

\address{Institut f\"ur Theoretische Physik, Westf\"alische
 Wilhelms-Universit\"at M\"unster, Wilhelm-Klemm-Stra\ss{}e 9,
 D-48149 M\"unster, Germany}

\ead{michael.klasen@uni-muenster.de}

\begin{abstract}
We demonstrate that not only the production of virtual photons decaying into
low-mass lepton pairs, but also the one of weak bosons at large transverse
momenta is dominated by quark-gluon scattering. Measurements of these processes
at the LHC can therefore provide useful constraints on the parton densities in
the proton, in particular the one of the gluon, and their nuclear modifications.
\end{abstract}

\vspace*{-130mm}
\begin{flushright}
MS-TP-14-05
\end{flushright}
\vspace*{119mm}

\section{Introduction}

\vspace*{3mm}

Parton distribution functions (PDFs) in protons and nuclei are traditionally
determined in deep-inelastic electron scattering. While the (valence)
quark densities can in this way well be determined, the gluon and sea quark
densities enter only at higher order in perturbation theory, so that they are
less well determined, in particular at very small and large momentum fractions
$x$. Therefore, constraints from other hard scattering processes like inclusive
hadron or jet production must be added in global analyses
\cite{Gao:2013xoa,Martin:2009iq,Ball:2011mu}. In nuclear
collisions, the free proton PDFs are modified by the presence of the
surrounding nucleons. From small to large $x$, shadowing, antishadowing
and EMC effects, Fermi motion and isospin modifications occur. These
effects are currently much less well determined than the PDFs in free protons
and are parametrised by various groups \cite{Eskola:2009uj,Hirai:2007sx,%
deFlorian:2011fp,Schienbein:2009kk}.

\vspace*{3mm}

Prompt photon production at sufficiently large transverse momentum $p_T$ is known
to be dominated by quark-gluon scattering, the so-called QCD Compton process,
rather than quark-antiquark scattering. It can therefore in principle
provide additional constraints on the gluon density and its nuclear modifications
\cite{Arleo:2007js,Arleo:2011gc}. However, fragmentation contributions
lead to additional uncertainties \cite{Bourhis:1997yu,Klasen:2013mga} and must
be reduced by applying isolation criteria \cite{Catani:2013oma}. These difficulties
do not occur with slightly virtual photons decaying into lepton pairs
with small invariant mass, which are still dominantly produced in the QCD
Compton process and can thus serve as a surrogate for prompt-photon
production \cite{Berger:1998ev,Berger:1999es}.
As we have recently shown and will describe in this presentation,
even $Z$ and $W$ boson production can serve to constrain the gluon PDFs in
the proton at large $x$, if the transverse momentum of the produced vector
boson is much larger than its mass \cite{Brandt:2013hoa}, while nuclear
modification effects can be constrained from low-mass lepton pair production
in proton-ion collisions at the LHC \cite{Brandt:2014vva}.

\section{Massive vector boson production at the LHC}

\vspace*{3mm}

\begin{figure}
\begin{center}
\includegraphics[width=0.64\textwidth,angle=-90]{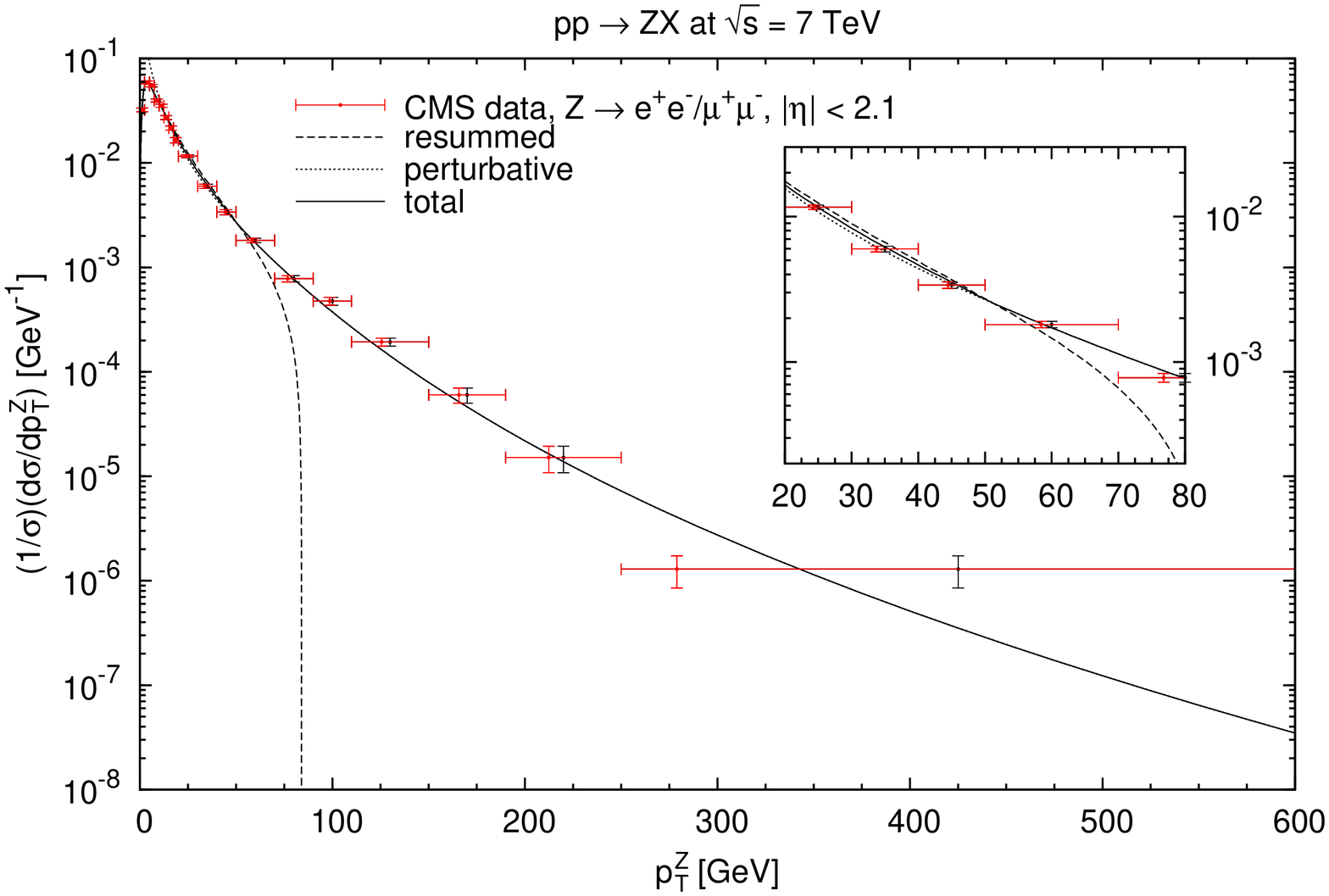}
\includegraphics[width=0.64\textwidth,angle=-90]{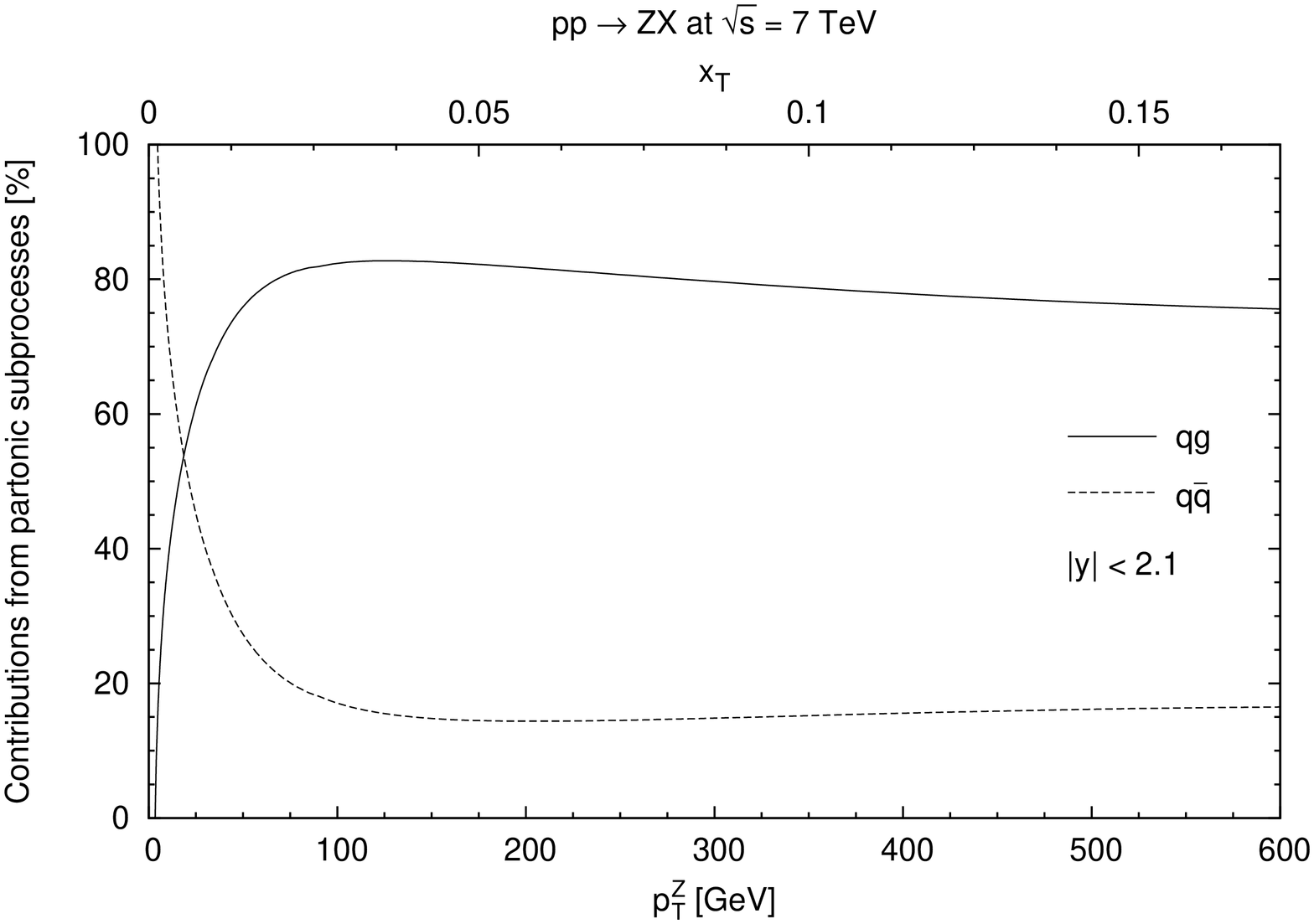}
\end{center}
\caption{\label{fig:1}Transverse-momentum distribution of $Z$ bosons produced
 in $pp$ collisions at the LHC, compared to recent CMS data (top), and its
 relative contributions from partonic subprocesses (bottom).}
\end{figure}

The production of $W$ and $Z$ bosons at the LHC with a centre-of-mass energy
of $\sqrt{s}=7$ TeV has recently been measured by the ATLAS \cite{Aad:2011fp}
and CMS \cite{Chatrchyan:2011wt} collaborations out to $p_T$ values of 600 GeV.
The corresponding distribution for $Z$ bosons is shown in Fig.\ \ref{fig:1} (top).
While at $p_T>65$ GeV our perturbative calculation at next-to-leading order (NLO)
is sufficient, this calculation diverges at low $p_T$ due to large logarithms
and must be resummed, which we do up to next-to-leading logarithmic accuracy.
As one can see, the theoretical predictions then agree very nicely with the 
data also at small $p_T$. Similar results have been obtained for the production
of $W$ bosons \cite{Brandt:2013hoa}.

\vspace*{3mm}

Intuitively, one might expect massive $Z$ and $W$ bosons to be produced dominantly
through the Drell-Yan process, i.e.\ in quark-antiquark annihilation, where a large
fraction of longitudinal momentum is transferred from the proton to the parton.
At the LHC, this is only true at small transverse momenta of the vector boson,
as can be seen from Fig.\ \ref{fig:1} (bottom). When $p_T> M_{W,Z}$, the mass
of the vector boson becomes less relevant, so that the QCD Compton process
eventually takes over as it does for real and virtual photon production.

\section{Parton densities in the proton}

\vspace*{3mm}

The values of $x$ probed by the current data sets, as estimated by $x_T=2p_T/\sqrt{s}$
at central rapidities, extend only to small and intermediate values of 0.01 ... 0.15,
where the quark and gluon densities are relatively well constrained.

\vspace*{3mm}

\begin{figure}
\begin{center}
\includegraphics[width=0.64\textwidth,angle=-90]{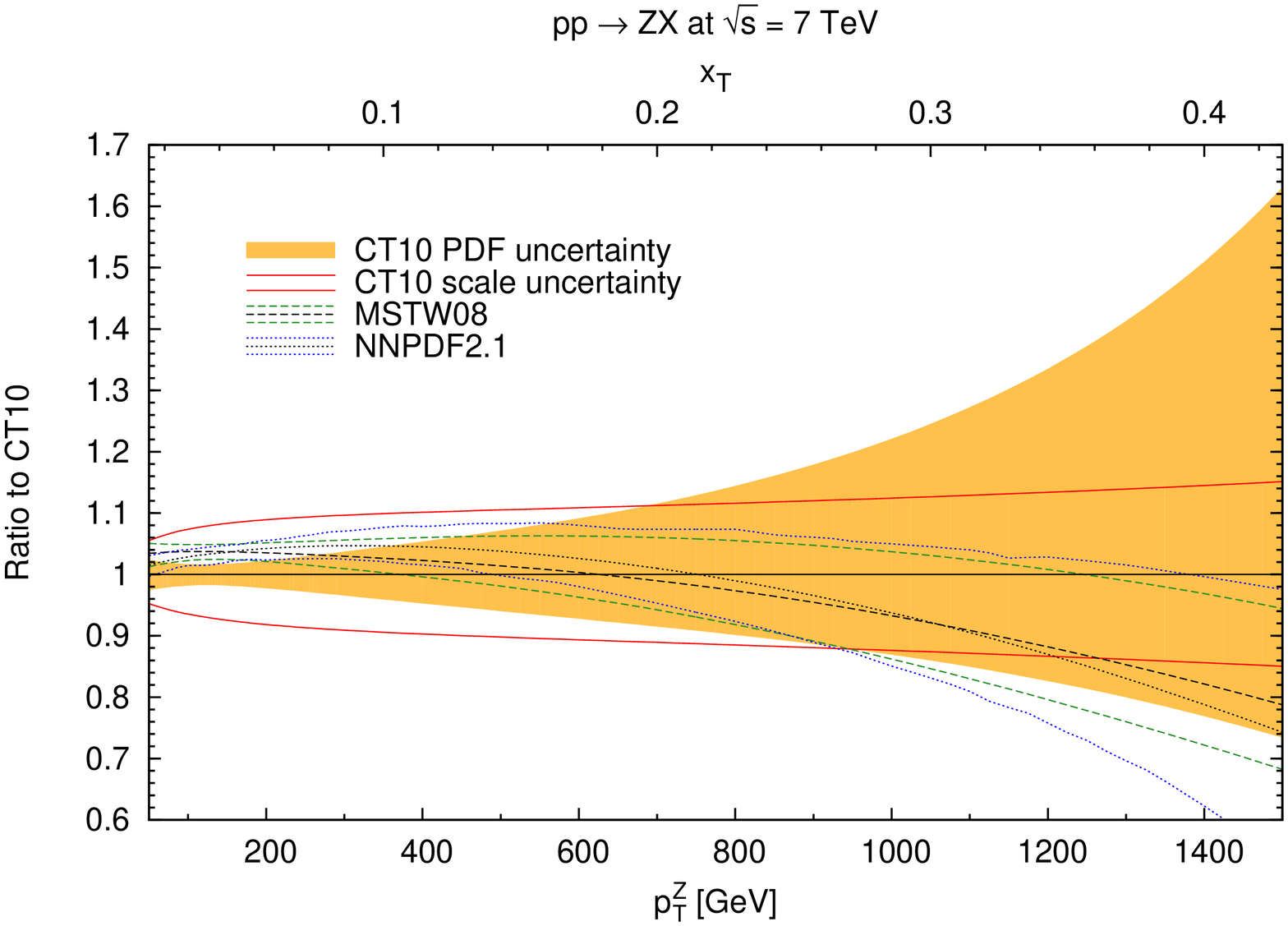}
\includegraphics[width=0.64\textwidth,angle=-90]{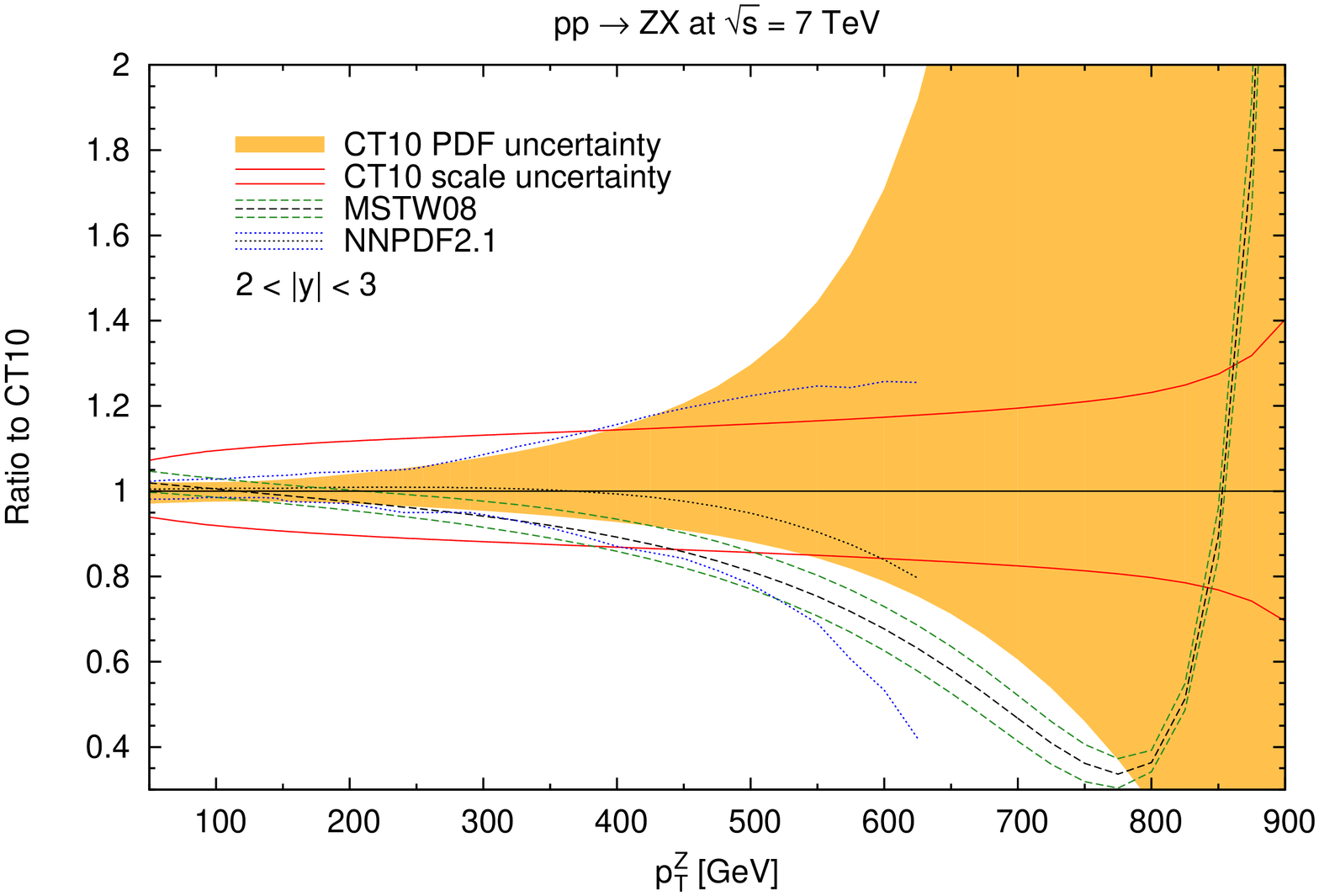}
\end{center}
\caption{\label{fig:2}Relative PDF und scale uncertainties of $p_T$-distributions for $Z$
 bosons produced in $pp$ collisions with $\sqrt{s}=7$ TeV at central (top) and forward
 (bottom) rapidity.}
\end{figure}

In Fig.\
\ref{fig:2} (top) we show that this changes dramatically as higher values of $p_T$ and
$x$ are reached. The PDF uncertainty as estimated by the CT10 error sets (yellow band)
then easily exceeds the theoretical uncertainty from scale variations at NLO (red lines).
In addition, other PDF parametrisations can fall outside the CT10 error band, indicating
that the latter has been underestimated due to the theoretical ansatz at the starting
scale or due to the selection of data sets entering the global analysis.
Reaching higher values of $p_T$ of course requires higher luminosity, which will however
soon become available at centre-of-mass energies of up to $\sqrt{s}=14$ TeV. At this
energy, the $x$-axes of our figures can simply be rescaled by a factor of two.

\vspace*{3mm}

Alternatively, large values of $x$ can be reached in the forward or backward rapidity
regions. As we show in Fig.\ \ref{fig:2} (bottom), the PDF uncertainties then exceed
those of the scale variations already at currently accessible $p_T$ values and can
easily reach 100\%.

\section{Nuclear modifications of parton densities}

\vspace*{3mm}

Nuclear modifications of parton densities can be determined in ratios of cross
sections of proton-ion over proton-proton collisions. These offer the advantage
that theoretical uncertainties from scale variations and free proton PDFs cancel
out to a large extent in the numerator and denominator, as do most experimental
systematic uncertainties.

\vspace*{3mm}

\begin{figure}
\begin{center}
\includegraphics[width=0.64\textwidth,angle=-90]{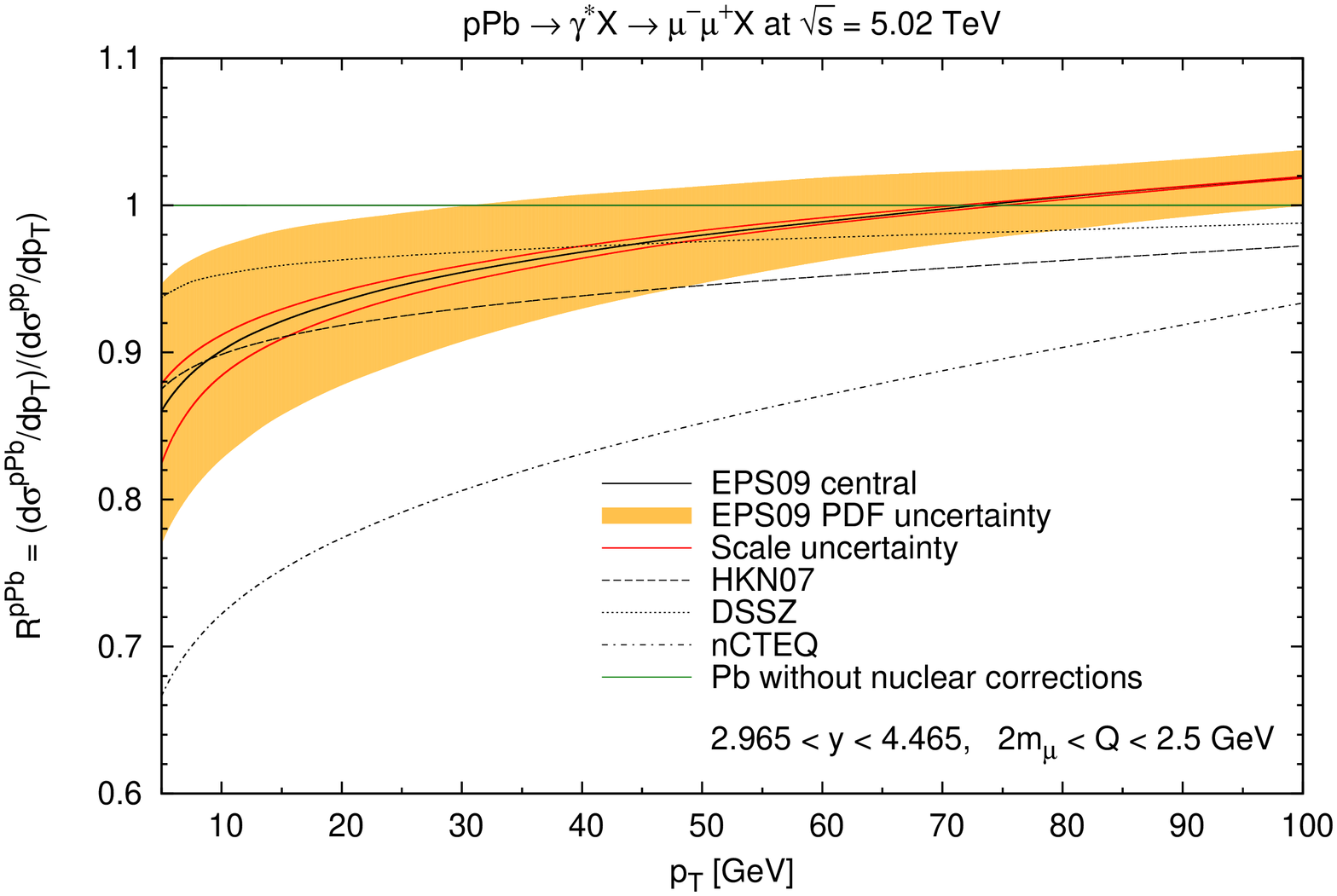}
\includegraphics[width=0.64\textwidth,angle=-90]{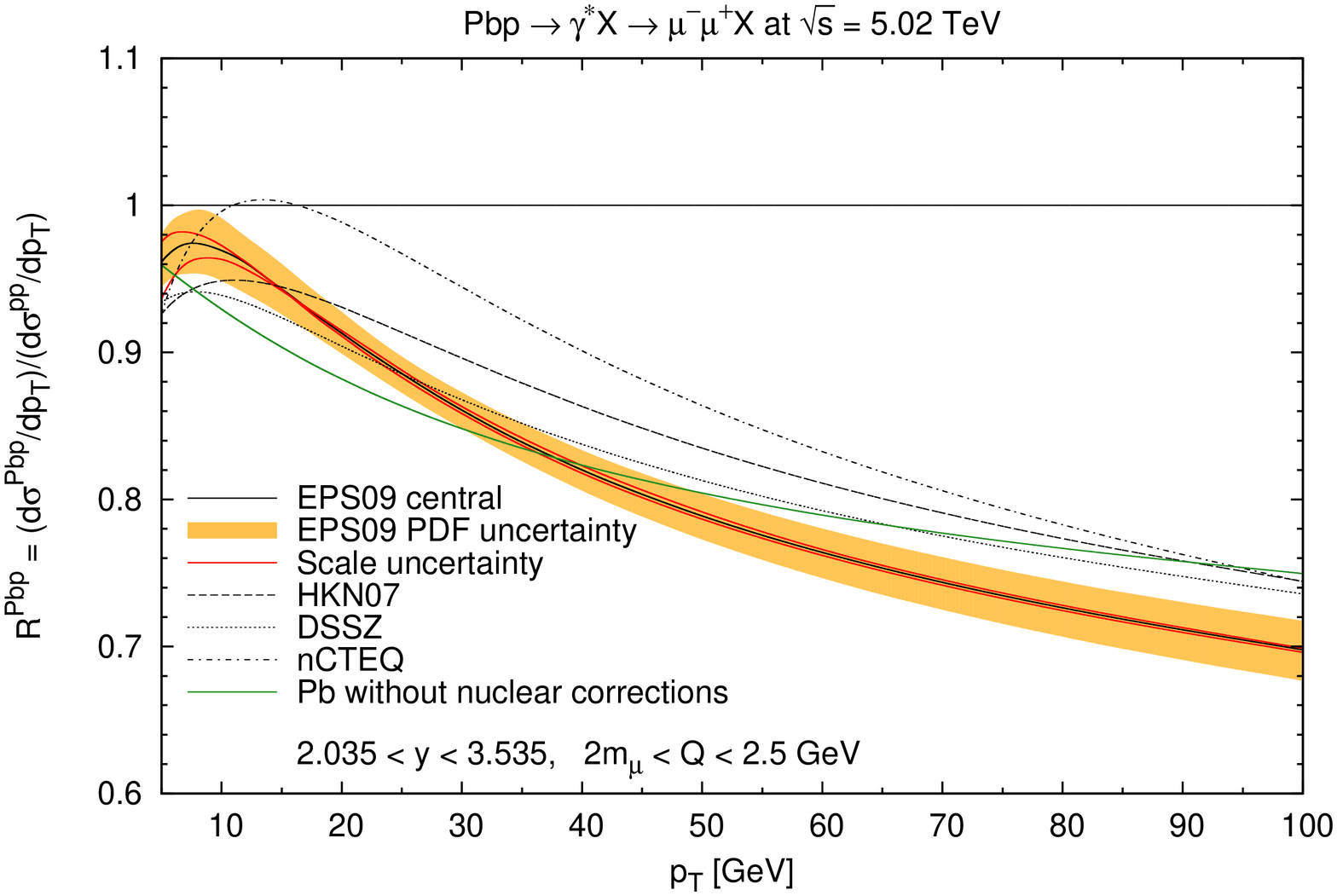}
\end{center}
\caption{\label{fig:3}Nuclear modification factors as a function of $p_T$ of
 low-mass muon pairs produced in proton-lead (top) and lead-proton (bottom) collisions
 with $\sqrt{s}=5.02$ TeV at forward rapidity.}
\end{figure}

In 2013, the LHC collided protons with lead ions at a centre-of-mass energy
of $\sqrt{s}=5.02$ TeV with beams circulating in both directions. Reference
data were also taken in proton-proton collisions at the same energy.
In Fig.\ \ref{fig:3} we show that the cross section ratios for low-mass muon
pairs produced in the forward direction and detected, e.g., with the ALICE muon
spectrometer, are very sensitive to nuclear modification effects.
If the protons circulate towards the muon spectrometer, the PDFs in the
lead ion are produced at relatively low values of $x$, i.e. in the shadowing
region (Fig.\ \ref{fig:3}, top). Only at relatively large values of
$p_T>70$ GeV the antishadowing region is reached, and the exact transition
point depends strongly on the nuclear PDF parametrisation. The uncertainty
in the shadowing effect is $\pm10$\% at low $p_T$ according to EPS09 and
reaches -20\% in the nCTEQ prediction, demonstrating again the strong bias
of the theoretical ansatz at the starting scale and of the selection of data
sets in the global analysis.

\vspace*{3mm}

When the lead ions
circulate towards the muon spectrometer (Fig.\ \ref{fig:3}, bottom), large
$x$ values are probed. There, one has to take into account the isospin
effect (green curve) and the EMC effect, which for EPS09 is predicted for
$p_T>40$ GeV and $x>0.4$. In the other parametrisations the antishadowing
region extends over almost the entire $p_T$ range shown.
Similar studies can be performed with electron-positron pairs in the central
detectors of ALICE, ATLAS and CMS. With the latter two general purpose
experiments, also much higher values of $p_T$ may eventually be reached.

\section{Conclusion}

\vspace*{3mm}

In conclusion, not only low-mass lepton pairs can serve as important probes of
the gluon density in the proton, in particular at large $x$, but also the
weak $W$ and $Z$ bosons can play this role despite their mass, provided that
their transverse momenta are sufficiently large for the QCD Compton process
to become dominant.
With low-mass lepton pairs,
nuclear modifications of parton densities can in principle be studied in
all relevant regimes, provided that sufficient beam time is allocated
to proton-ion runs at the LHC to reach luminosities at least in the pb range.
Measuring ratios of cross sections then allows for large cancellations of
theoretical and experimental uncertainties, leading to considerable improvements
of our current knowledge of parton distributions in bound protons.

\ack

\vspace*{3mm}

The author thanks the organisers of this conference for the kind invitation
to this very stimulating workshop and M.\ Brandt, C.\ Klein-B\"osing, F.\
K\"onig and J.\ Wessels for their collaboration.

\end{document}